# Raman spectral shift versus strain and composition in GeSn layers with: 6 to 15% Sn contents

A. Gassenq[a*], L. Milord[b], J. Aubin[b], N. Pauc[a], K.Guilloy[a], J. Rothman[b], D. Rouchon[b], A. Chelnokov[b], J.M. Hartmann[b], V. Reboud[b], V. Calvo[a]

[1]Univ. Grenoble Alpes, CEA-INAC, 17 rue des Martyrs, 38000, Grenoble, France

[2]Univ. Grenoble Alpes, CEA-LETI, Minatech, 17 rue des Martyrs, 38000, Grenoble, France

**ABSTRACT**

GeSn alloys are the subject of intense research activities as these group IV semiconductors present direct bandgap behaviors for high Sn contents. Today, the control of strain becomes an important challenge to improve GeSn devices. Strain micro-measurements are usually performed by Raman spectroscopy. However, different relationships linking the Raman spectral shifts to the built-in strain can be found in the literature. They were deduced from studies on low Sn content GeSn layers (i.e. $x_{Sn}<8\%$) or on GeSiSn layers. In this work, we have calibrated the GeSn Raman relationship for really high Sn content GeSn binaries ($6<x_{Sn}<15\%$). We have used fully strained GeSn layers and fully relaxed GeSn under-etched microstructures to clearly differentiate the contributions of strain and chemical composition on the Ge-Ge Raman spectral shift. We have shown that the GeSn Raman-strain coefficient for high Sn contents is higher compared to that for pure Ge.

* Corresponding author. E-mail address: alban.gassenq@cea.fr

GeSn is a very interesting group IV material which presents a direct bandgap (predicted since the eighties[1,2]) that could be most interesting for photonics[3] or microelectronics applications[4]. Thanks to progresses in Chemical Vapor Deposition (CVD)[5], laser operation has been demonstrated in 2015[3] and confirmed in 2016[6,7], using high Sn content $Ge_{1-x}Sn_x$ layers ($x_{Sn}>8\%$). Different possibilities are currently investigated to improve such devices. It has been predicted that the tensile strain would allow to reduce the Sn content needed to have a direct bandgap[8–12]. Furthermore, compressively-strained GeSn layers are promising for transistors[13] or Mid Infra-Red photodetectors[14,15] applications. Therefore, the control of strain in GeSn devices is now essential to develop and improve actual GeSn components. Raman spectroscopy is a suitable technics for strain characterization at the micron-scale. However, many different GeSn Raman-strain relationships can be found in the literature. It is worthwhile noting that they were determined from studies on low Sn contents binaries (<8%)[16–24] or GeSiSn ternary alloys[25]. Therefore, differences between them need to be better understood and an upgraded formula for GeSn with higher Sn contents has to be established.

In this work we have measured the Raman-strain relationship in GeSn layers with high Sn content, between 6 and 15% Sn. The GeSn layers were grown on Ge strain-relaxed buffers on Si substrates. X-Ray Diffraction (XRD), photoluminescence (PL) and Raman spectroscopy showed that layers were of high crystalline and optical quality and that the strain relaxation increased with the layer thicknesses, as expected. To dissociate strain ($\varepsilon$) and chemical composition ($x_{Sn}$) contributions to the Raman spectral shift ($\Delta\omega$) associated with GeSn layers, we have measured $\Delta\omega$ in (i) thin, fully compressively strained layers (i.e. pseudomorphic GeSn on Ge) and (ii) under-etched, fully relaxed microstructures fabricated from such pseudomorphic layers. We have found a good agreement with the literature for the Raman-Sn content coefficient. However, we have shown that the Raman-strain coefficient increases for high Sn contents compared to that for pure Ge.

GeSn layers were grown in a 200-mm Epi Centura 5200 Reduced Pressure CVD tool[26–28]. 1.3µm Ge Strain Relaxed Buffers (SRB) were grown on Si(001) substrates with a small residual tensile strain of around 0.2%[29–33]. Two parameters were changed: the Sn content and the GeSn layer thickness controlled by the growth conditions (i.e. growth temperature and duration). The Sn content in the GeSn layer and the growth rate have been calibrated by XRD[27,28]. **Figure 1-a** shows high resolution conventional ω-2θ scans around the (004) XRD order performed on 30 nm thick GeSn layers grown at different temperatures. The intense peaks associated with the Si substrate and the Ge buffer are present at high angles in each profile. As expected for pseudomorphic layers, we have at lower angles intense GeSn layer peaks with thickness fringes indicating a very good quality of the GeSn layers with abrupt GeSn/Ge interfaces. Thanks to the Takagi-Taupin's dynamical diffraction theory[34] we have measured the out-of-plane lattice parameters and calculated the corresponding lattice parameter[35,36]. By taking into account the positive deviation from a straightforward interpolation between the lattice parameter of Ge and Sn[35,37], we have deduced the Sn concentration $x_{Sn}$ in such layers. For the same growth pressure and precursor flows, Sn contents can be tuned from 6% up to 15% by changing the growth temperature in the 300-350°C range [26,28]. For thicker layers, Reciprocal Space Mappings (RSM) around the (224) asymmetric diffraction order have been performed. **Figure 1-c** shows typical RSMs for various thickness $x_{Sn}$~10 % GeSn layers. For thinner samples, the GeSn layer peak is on the pseudomorphic dashed red line. When the thickness increases, the GeSn diffracted peaks shift toward the full relaxation doted green line and becomes broader due to mosaicity[38]. RSM measurements shows that our 30 and 60 nm layers are pseudomorphic with very low strain relaxation values (R<1.5%) because layer thicknesses are below the critical thickness for plastic relaxation[35,39].

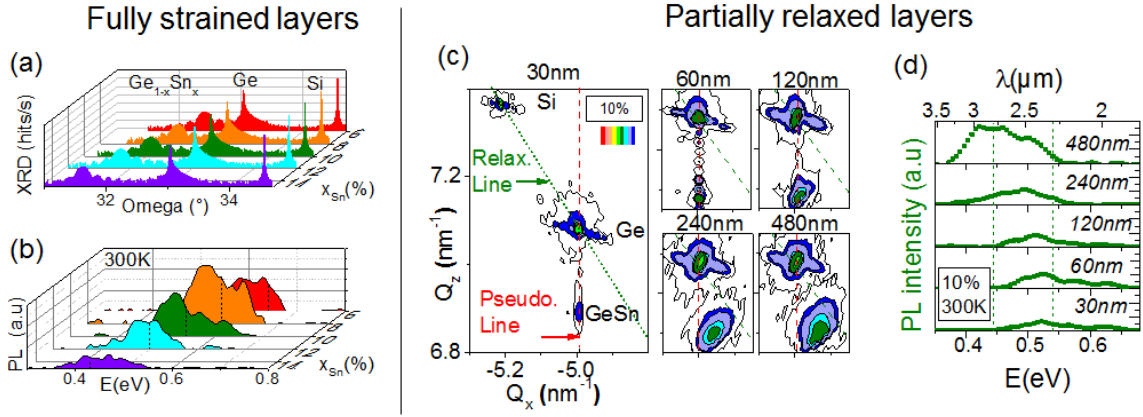

**Figure 1:** (a) XRD Omega-2Theta scans for different Sn contents thin, fully strained layers; (b) PL spectra associated with those fully strained layers; (c) XRD reciprocal space mapping for different thickness, partly relaxed $Ge_{90}Sn_{10}$ layers; (d) PL spectra associated with those different thickness $Ge_{90}Sn_{10}$ layers.

PL spectra associated with those thin, fully strained GeSn layers are presented in **Figure 1-b**. PL measurements were performed using a pulsed laser emitting at 1047 nm (10 kHz, 10 ns) at room temperature. The pump laser was focused on a 20 µm diameter spot with a 1 mW average power. The pump power was low enough to avoid heating. Light was sent to a home-built Fourier Transform Infra-Red spectrometer[40] with a mercury cadmium telluride avalanche photodetector[41]. As expected, a clear red shift is observed as the Sn content in those layers increases from 6% up to 15%.[42–48] **Figure 1-d** shows PL spectra for different thickness $Ge_{90}Sn_{10}$ layers. A red shift is detected as the layer thicknesses increases (and the layer plastically relaxes). The theoretical Gamma bandgap values have been added as vertical dashed lines in Figure 1-d. The relaxed bandgap is evaluated to be around 0.45eV[12] and the pseudomorphic bandgap around 0.55 eV[28]. We clearly see that the spectrum associated with the pseudomorphic layer (i.e the 30 nm thick one) is centered on the strained value (0.55eV) while the spectrum of the partially relaxed layer (i.e with 480 nm) is centered on the relaxed bandgap value (0.45eV). XRD and PL measurements performed on our layers thus show that our layers are of high crystalline and optical quality, with a tunable amount of Sn from 6 to 15% and a degree of strain relaxation which, as expected, increases with the layer thickness.

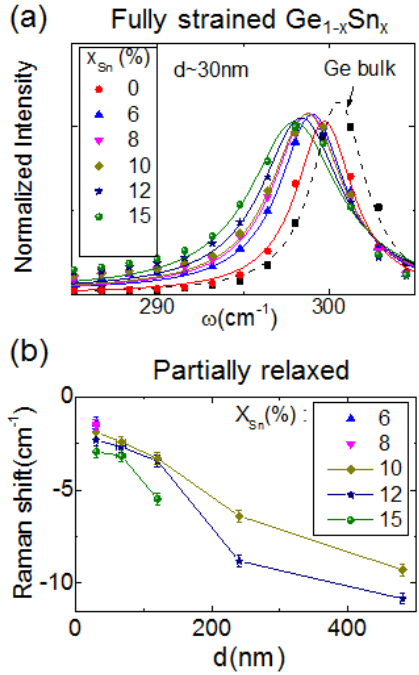

**Figure** 2: a) Measured Raman shifts for 30 nm thick, various Sn content GeSn layers; b) Measured Raman spectral shift for the Ge-Ge mode for different layer thicknesses d and compositions $x_{Sn}$.

Raman spectroscopy measurements have been performed on those layers. A Renishaw InVia Raman spectrometer was used with a 532 nm incident laser[49,50] (corresponding to a penetration depth of ~20 nm in Ge[51]) focused on a 0.7 µm-diameter spot. The Raman shift was measured by fitting the spectra with Lorentzian functions. A bulk Ge substrate was used as a reference for 0 % deformation. **Figure 2-a** shows the Raman spectra for the Ge-Ge mode[25] for 30 nm thick, various Sn contents fully strained GeSn layers. The Raman spectral shift increases with the amount of Sn. Note that the 0% Sn content data shows a Ge spectral shift with 0.2% of tensile strain which correspond to the typical value for Ge SRBs grown on Si[30–33]. **Figure 2-b** shows the influence of layer thicknesses on the Raman spectral shift for different Sn concentrations. The spectral shift uncertainties (±0.1cm$^{-1}$) are attributed to the spectral resolution of the spectrometer. The Raman spectral shift increases with the thickness due to strain relaxation, as expected[19–25]. Therefore, we have selected the thin 30 nm pseudo-morphic layers for measuring the GeSn Raman relationship.

$$\Delta\omega = a \times x_{Sn} + b \times \varepsilon_{//} \quad (1)$$

$$\Delta\omega_{pseudo} = a \times x_{Sn} + 0.147 \times b \times x_{Sn} \quad (2)$$

$$\Delta\omega_{pseudo} = c \times x_{Sn} \quad (3)$$

The Raman-strain relationships are functions of material composition and strain configuration.[52,53] For GeSn, the Raman spectral shift ($\Delta\omega$) is linked to the in-plane strain $\varepsilon_{//}$ and the Sn content $x_{Sn}$ by Equation 1,[16–25] $a$ being the Raman-Sn content coefficient and $b$ the Raman-strain shift coefficient. For pseudomorphic GeSn layers grown on Ge, the in-plane strain is proportional to the lattice mismatch, i.e. $0.147 x_{Sn}$ [20,36] (Equation 2), generating a spectral shift proportional to $x_{Sn}$ (Equation 3). **Table I** gives the reported Raman coefficients *a, b* and *c* available in the literature for GeSn alloys[16–25,54]. Note that the *c* coefficient in ref.[20] does not really apply for pseudomorphic layers (d~250nm) and that the pioneering work[16] is out of range due to an "inadequacy strain correction"[17] or "considerable disorders in their materials"[20].

**TABLE I:** Summary of the different Raman-Sn content-strain coefficients in the literature[16–23,25,54]

| a | b | c | $x_{Sn}$ | Ref. | Method |
|---|---|---|---|---|---|
| cm$^{-1}$ | cm$^{-1}$ | cm$^{-1}$ | % | | |
| -140 | 64 | | 0-22 | Lopez 1998[16] | GeSn on Ge |
| -66 | | | 0-20 | Li 2004[17] | GeSn on Ge |
| -75 | | | 0-18 | DaCosta 2007[18] | GeSn on Si |
| -82 | 563 | | 4-8 | Lin 2011[19] | GeSn on InGaAs |
| -95 | | -31 | 0-8 | Su 2011[20] | GeSn on Ge or Si |
| -93 | 415 | | 2-12 | Fournier 2013[25] | GeSiSn |
| -83 | 375 | -23 | 0-8 | Cheng 2013[21] | Relaxed and strained GeSn on Ge |
| -78 | 299 | | 0-8 | Chang 2015[23] | Annealed GeSn |
| -78 | 399 | | 0-3 | Takeuchi 2016[54] | GeSn in oil |
| -88 | 521 | -13 | 6-15 | This work | Relaxed and strained GeSn on Ge |

In order to measure the GeSn Raman relationships, we have first extracted the *a* coefficient using fully relaxed GeSn micro-structures fabricated from our pseudomorphic layers (i.e. with $\varepsilon_{//}$=0% in Equation 1). The *b* and *c* coefficients were then measured on 30 nm thick pseudomorphic layers (Equations 2 and 3). Such a method has already been used to measure the Raman-strain coefficient, for lower Sn contents ($x_{Sn}$<8%).[21] **Figure 3-a** shows a schematics of the fully relaxed microstructures that we have fabricated. Fully strained GeSn layers were first of all patterned using ultra violet lithography followed by an anisotropic dry etching with $Cl_2/N_2$ gasses. The GeSn layers were then under-etched using a selective dry etching recipe based on $CF_4$.[55,56] **Figure 3-b** and **3-c** are tilted Scanning Electron Microscopy (SEM) views of the fabricated microstructures. Two designs have been tested: micro-disks (Figure 3-b) and micro-edges (Figure 3-c). A fraction of the GeSn layer is suspended over more than 1 µm. In that region, the strain elastically relaxes, giving the observed undulations. **Figure 3-d** shows Raman spectra for Fig. 3-c $Ge_{90}Sn_{10}$ microstructure. The laser intensity was low enough (~4 µW) to avoid any heating effects.[57] The presented spectra were measured on the top surface of the GeSn waveguide (position A in Fig. 3-c) and in the suspended region (position B in Fig. 3-c). A very large Raman shift is detected in the under etched layer. Line scan measurements along the X axis presented in Figure 3-c were also performed (inset of the Figure 3-d). Two distinct areas with an abrupt transition are detected in the Raman spectral shift which is a clear indication that strain is fully relaxed in the suspended region.

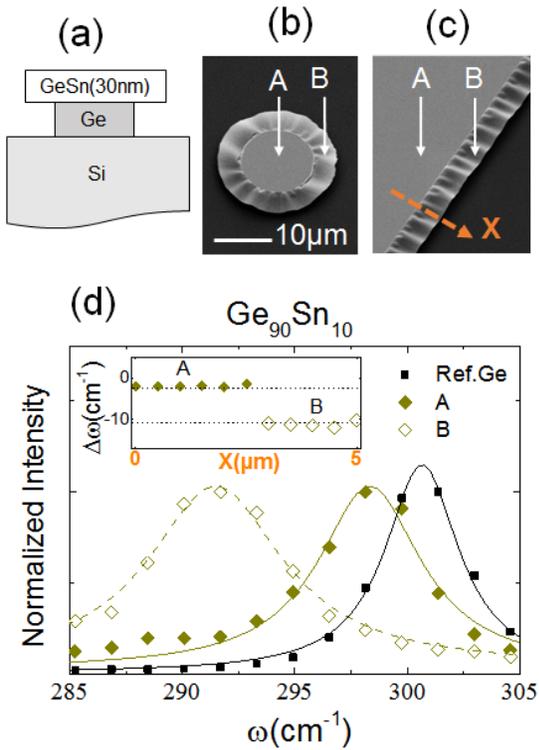

**Figure 3**: (a) Under-etched GeSn/Ge microstructures; tilted SEM imaging of fabricated devices: under-etched (b) micro-disk and (c) micro-edge; (d) Raman spectra measured in pseudomorphic or under-etched regions compared to that for a bulk Ge substrate (the inset shows a Raman shift line scan measurement performed along the X axis presented in the Figure 3-c).

Raman spectra shifts for fully relaxed microstructures and fully strained GeSn layers are shown as dots in **Figure 4**, as a function of the Sn content. A good agreement is found between the under-etched micro-disks and micro-edges (**Figure 4-a**). The corresponding in-plane strain for pseudomorphic layers (**Figure 4-b**) has been added in the top axis corrected by the typical -0.2% value coming from the Ge buffer[30–33]. In addition, all the reported $a$ and $b$ coefficients provided in Table I[16–25] have been added as dashed lines and our data linear fits are plotted as solid lines.

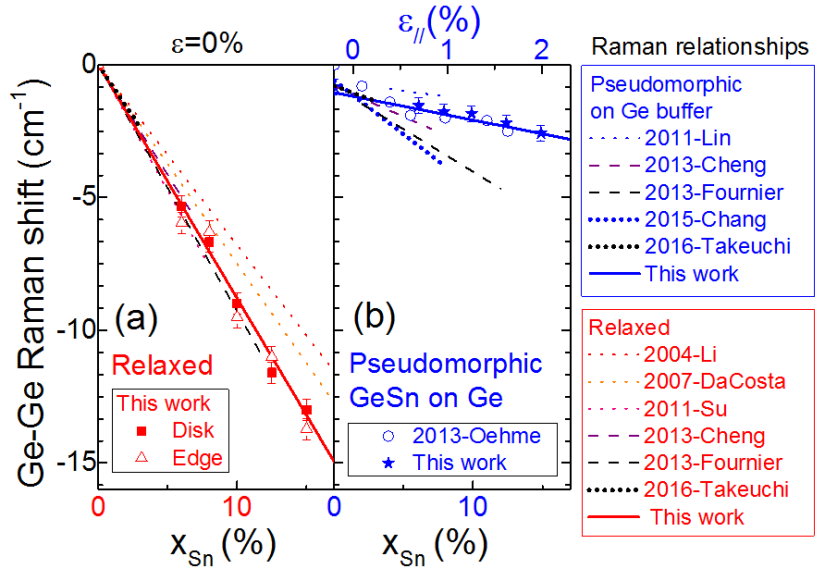

**Figure 4**: State of the art for the Raman shift as a function of Sn content for relaxed[17–21,25,54] and strained layers[19,21,23,25,54] compared to experimental data from[58] and our work.

By fitting our data, we found a=-88($\pm$3) cm$^{-1}$, b=521($\pm$15) cm$^{-1}$ and c=-13($\pm$3) cm$^{-1}$. The indicated uncertainties correspond to fit deviations. As far as the Raman-Sn content coefficient is concerned (i.e. a=-88 cm$^{-1}$), a good agreement is found with the most recent works.[19–21,25] The difference with the oldest works[17,18] is likely due to an incorrect full relaxation assumption[17] or to a strain correction using the Ge strain shift coefficient[18].

Our Raman-strain coefficient (i.e. b=521 cm$^{-1}$) is by contrast higher than previously reported coefficients (Table I) excepted for[19] which is the only study on thin GeSn layers grown on InGaAs SRBs. Differences with other works can be attributed to the use of thick annealed layers[23] or to the layer compositions[21,25,54]. For thick annealed layers (i.e. 160 nm at 8%)[23], strain relaxation[39,59,60] or Sn diffusion[61–63] can lead to an underestimation of the Raman-strain coefficient[23]. For the other reported GeSn Raman-strain coefficients (b = 399[24] and 375 cm$^{-1}$,[21] obtained for $x_{Sn}$<8%), they are expected to be closer to the Ge one (b~400m$^{-1}$)[52,64–67] compared to our coefficient coming from higher Sn content layers (b=521cm$^{-1}$ for 6<$x_{Sn}$<15%). Indeed, the GeSn Raman-strain coefficient is expected to be Sn content dependent. It however stays relatively constant in our Sn content range (i.e. 6<$x_{Sn}$<15%), given our measurement uncertainties. Since the only Raman-strain coefficient reported for high Sn contents (e.g. up to 12%) was measured in GeSiSn ternary layers using RSM strain extraction, it cannot really be confronted to our data. The only other work which can be compared with our work is ref[58]. The authors of that paper studied thin, high Sn content GeSn layers (up to 12.5%). They mentioned a strain shift coefficient which increased with the Sn content. Such data have been added in Figure 4 (blue circle). They are in very good agreement with our experiments. Therefore, our GeSn Raman relationship is actually the only one which has been determined using high Sn content (6<x<15%) GeSn fully strained layers and fully relaxed micro-structures. Such a relationship enables a more precise characterization of high Sn content GeSn devices with Raman spectroscopy.

To sum up, we have grown high quality GeSn layers on Ge strain-relaxed buffers. XRD, PL and Raman spectroscopy evidenced a plastic strain relaxation which increased with the layer thickness. We have then used the pseudomorphic GeSn layers to calibrate the Raman-strain versus Sn composition relationship using fully relaxed under-etched micro-structures and fully compressively-strained layers. By comparing our measured equation to literature, we have shown that the GeSn Raman-strain coefficient for high Sn contents (6<$x_{Sn}$<15%) is higher compared to that of pure Ge. This work provides a better understanding of GeSn material for photonics and micro-electronics applications.

**Acknowledgements**

The authors would like to thank the Platforme de Technologie Amont of the CEA Grenoble for the clean room facilities. This work was supported by the CEA DSM-DRT Phare Photonics and IBEA Nanoscience projects